\documentclass[letterpaper]{article} 
\usepackage{aaai2027}  

\usepackage[hyphens]{url}  
\usepackage{graphicx} 
\usepackage{natbib}  
\usepackage{caption} 
\usepackage{algorithm}
\usepackage{algorithmic}
\usepackage{amsmath}
\usepackage{amssymb}
\usepackage{subcaption}
\usepackage{newfloat}
\usepackage{listings}
\DeclareCaptionStyle{ruled}{labelfont=normalfont,labelsep=colon,strut=off} 
\floatstyle{ruled}
\newfloat{listing}{tb}{lst}{}
\floatname{listing}{Listing}

\usepackage{booktabs}

\title{Direction-adaptive Mamba: Spatial-Frequency Dual-Domain Collaborative Learning for PolSAR Image Classification}

\author{
    Junfei Shi,
    Yu Cheng,
    Haojia Zhang,
    Wenqiang Hua,
    Junhuai Li,
    Maoguo Gong
}
\affiliations{

}

\begin{document}

\maketitle

\begin{abstract}
Deep learning dominates polarimetric synthetic aperture radar (PolSAR) image classification, with Mamba architectures serving as favorable backbones due to linear complexity and strong global modeling capacity. However, existing PolSAR Mamba methods have two critical flaws: pure spatial processing discards fine-grained edges and textures, and fixed scanning patterns fail to model direction-variant anisotropic scattering and weak boundaries essential for PolSAR physical analysis. This work proposes DA-Mamba, a direction-adaptive Mamba framework with dual-domain collaborative learning for PolSAR classification. Equipped with an edge-aligned direction-adaptive scanning scheme, DA-Mamba captures long-range spatial dependencies and accurate boundary details. It adopts the Non-Subsampled Contourlet Transform (NSCT) to separate PolSAR data into low-frequency global components and multi-directional high-frequency subbands, extracting anisotropic structural features from high-frequency information while preserving global context via low-frequency branches. A dual-domain collaborative learning module further integrates spatial scattering and frequency-domain representations to strengthen feature discriminability. Evaluated on three real-world PolSAR datasets, DA-Mamba surpasses state-of-the-art methods, verifying the efficacy of the proposed adaptive scanning and dual-domain fusion designs. Code will be publicly available.
\end{abstract}


\section{Introduction}
Polarimetric Synthetic Aperture Radar (PolSAR) image classification plays a critical role in diverse Earth observation and remote sensing applications, including land cover mapping\cite {RN1}, agricultural monitoring\cite {RN2}, urban planning and disaster assessment\cite {RN4}. Benefiting from abundant scattering matrix measurements, PolSAR data enables more comprehensive physical characterization of ground targets than single-polarization SAR. Traditional PolSAR classification methods rely on handcrafted features derived from scattering mechanisms\cite {RN11} and statistical distributions\cite {RN12}, which fail to capture the complex, hierarchical scattering patterns prevalent in heterogeneous terrain scenarios.

Recently, deep learning has emerged as the dominant paradigm for PolSAR image interpretation, yielding state-of-the-art performance in terrain classification and target detection. Convolutional Neural Networks (CNNs)\cite {RN14,RN15} are widely prevalent in this field for their exceptional local feature extraction ability, yet they are inherently constrained by fixed local receptive fields and fail to model long-range global context. By contrast, Vision Transformers (ViTs)\cite {RN16,RN17} attain full global receptive fields via self-attention mechanisms. However, ViTs incur input-dependent quadratic computational complexity, which severely hinders their deployment on high-resolution PolSAR imagery.

To address the aforementioned limitations, the Mamba state-space architecture has emerged as a promising alternative, enabling linear-complexity global context modeling with favorable tradeoffs between performance and computational overhead\cite {RN5,RN6}. Nonetheless, existing PolSAR Mamba variants\cite {RN18} focus solely on spatial-domain feature learning. This spatial-only design neglects fine-grained edges and heterogeneous structural cues critical for differentiating PolSAR scattering mechanisms, frequently causing boundary blurring and edge misclassification. While frequency-domain methods can remedy this defect by extracting discontinuity-emphasized detail features, naive frequency-domain 2D Mamba deployment ignores directional priors. This deficiency prevents capturing the inherent anisotropy of PolSAR edges, leading to inaccurate and inconsistent edge delineation.
\begin{figure}[htbp]
    \includegraphics[scale=0.6]{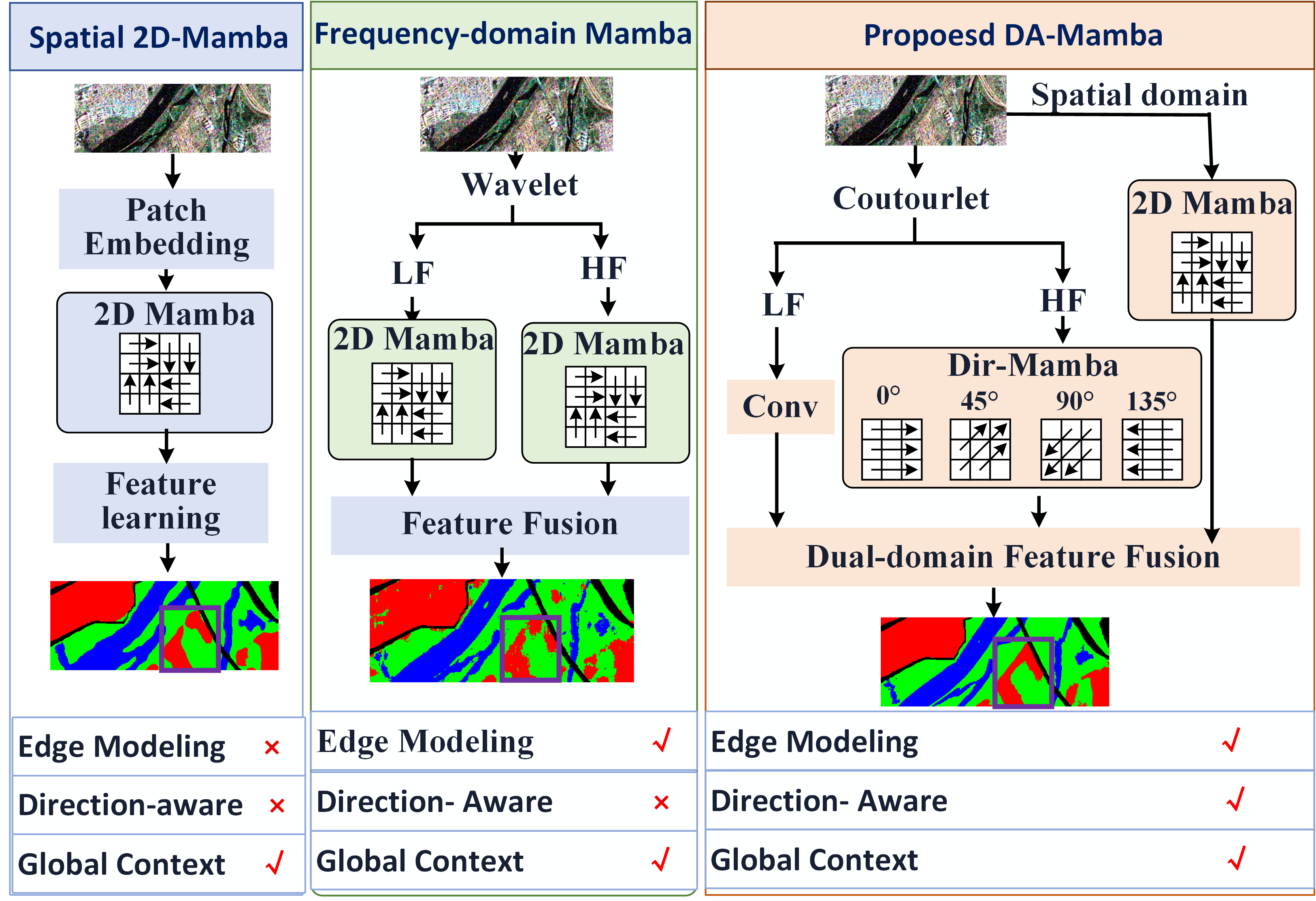} 
    \caption{Comparison of conventional and proposed methods.}
    \label{fig:comPareFig} 
\end{figure}

Fig.~\ref {fig:comPareFig} illustrates the architectures of spatial 2D-Mamba, frequency-domain Mamba, and our proposed dual-domain direction-adaptive Mamba (DA-Mamba). Compared with the two baseline Mamba variants, our method excels at extracting fine structural features in the frequency domain while capturing long-range global context in the spatial domain. This dual-domain collaborative learning paradigm yields more discriminative feature embeddings and boosts final classification performance. Experiments on the Xi’an dataset further verify the efficacy of our DA-Mamba module and the overall framework.



Consequently, existing Mamba methods still face two fundamental limitations: (1) they neglect critical edge details during feature learning, leading to pronounced edge confusion and misclassification at object boundaries; and (2) they rely solely on either the spatial or frequency domain in isolation, thus failing to holistically integrate local structural details with global contextual cues, resulting in suboptimal feature representations.

To tackle the above drawbacks, we propose DA-Mamba-DDCL, a dual-domain collaborative learning framework integrated with a direction-adaptive Mamba tailored for PolSAR feature extraction across spatial and frequency domains. Distinct from standard fixed-scanning Mamba, our edge-aligned adaptive scanning captures long-range structural dependencies and precisely outlines object boundaries. We first adopt the Non-Subsampled Contourlet Transform (NSCT) to split PolSAR images into low-frequency global components and multi-orientation high-frequency subbands. The proposed direction-adaptive Mamba(DA-Mamba) processes high-frequency subbands along orientation-aware trajectories to model anisotropic edges, while low-frequency branches retain holistic scene context. Additionally, our dual-domain architecture leverages a spatial 2D-Mamba to extract scattering signatures from raw PolSAR data. Features from the two domains are fused to support classification. The core contributions of this paper are outlined below:

1) We propose a novel direction-adaptive Mamba that dynamically scans along edge orientations, effectively capturing direction-sensitive structural information and alleviating edge confusion in PolSAR classification.

2) We design a dual-domain spatial-frequency collaborative learning framework that holistically integrates spatial scattering context with structural features of the frequency-domain, achieving more discriminative and robust representations.

3) Extensive experiments demonstrate that our DA-Mamba-DDCL framework achieves superior classification performance compared to state-of-the-art methods, validating the effectiveness of both the direction-adaptive scanning strategy and the dual-domain collaboration.

\section{Related Work}

\subsection{Deep Learning-Based Methods}
Driven by the rapid advancement of deep learning, PolSAR image classification has evolved from handcrafted feature engineering\cite {RN26,RN27,RN28} to end-to-end data-driven feature learning. Early mainstream studies adopted Convolutional Neural Networks (CNNs) to extract discriminative spatial and polarimetric PolSAR features\cite {RN33,RN35,RN37}. For instance, Zhou et al. first deployed CNNs for six-dimensional PolSAR data processing\cite {RN33}; Shang et al. developed a dual-branch CNN to boost feature representation capability\cite {RN35}; Zhang et al. further exploited 3D convolution to jointly model polarimetric properties and local spatial correlations\cite {RN36}.
Nevertheless, CNNs suffer from limited local receptive fields and fail to capture long-range spatial dependencies and global context. Attention and Transformer architectures have been adopted for PolSAR classification to remedy this issue\cite{RN38,RN39}, yet they incur heavy computation and memory overhead, particularly for high-dimensional multi-channel PolSAR data.

\subsection{Frequency-Domain Methods}

Frequency-domain multi-scale decomposition enriches PolSAR feature representations by isolating low-frequency semantics from high-frequency textures while suppressing speckle noise. Notable methods include LC-PSENet \cite{RN45}, which fuses low-frequency and contour subbands; Complex Contourlet-CNN \cite{RN41}, combining multi-scale contourlet features with complex convolutions; and C2N2 \cite{RN42}, which enhances contourlet-domain modeling via complex-valued learning. Wavelet-based approaches have also proven effective, such as using 3D discrete wavelet transforms \cite{RN43} and combining wavelet convolutions with hierarchical spatial transformers \cite{RN44}. However, these methods overlook directional correlations across high-frequency subbands, failing to fully capture the anisotropic scattering and directional textures intrinsic to PolSAR data.

\subsection{State Space Model}
State Space Models (SSMs) have recently emerged as an effective framework for long-range sequence modeling. S4 \cite{RN7} first introduced structured state space representations to efficiently capture long-range dependencies while maintaining linear complexity. Building upon S4, S5 \cite{RN6} further simplified the state-space formulation and improved computational efficiency through a multi-input multi-output design. More recently, Mamba \cite{RN5} incorporated selective state update mechanisms and hardware-aware optimization strategies into the SSM framework, significantly enhancing modeling capability and scalability. Owing to these advantages, Mamba has demonstrated promising performance across a variety of remote sensing tasks, including hyperspectral image classification \cite{RN19,RN20,RN21}, remote sensing scene classification \cite{RN22,RN23}, and semantic segmentation \cite{RN24,RN25}. Motivated by these successes, we investigate the application of Mamba to PolSAR image classification.

\section{Proposed method}

\begin{figure*} 
    \centering 
    \includegraphics[scale=0.9]{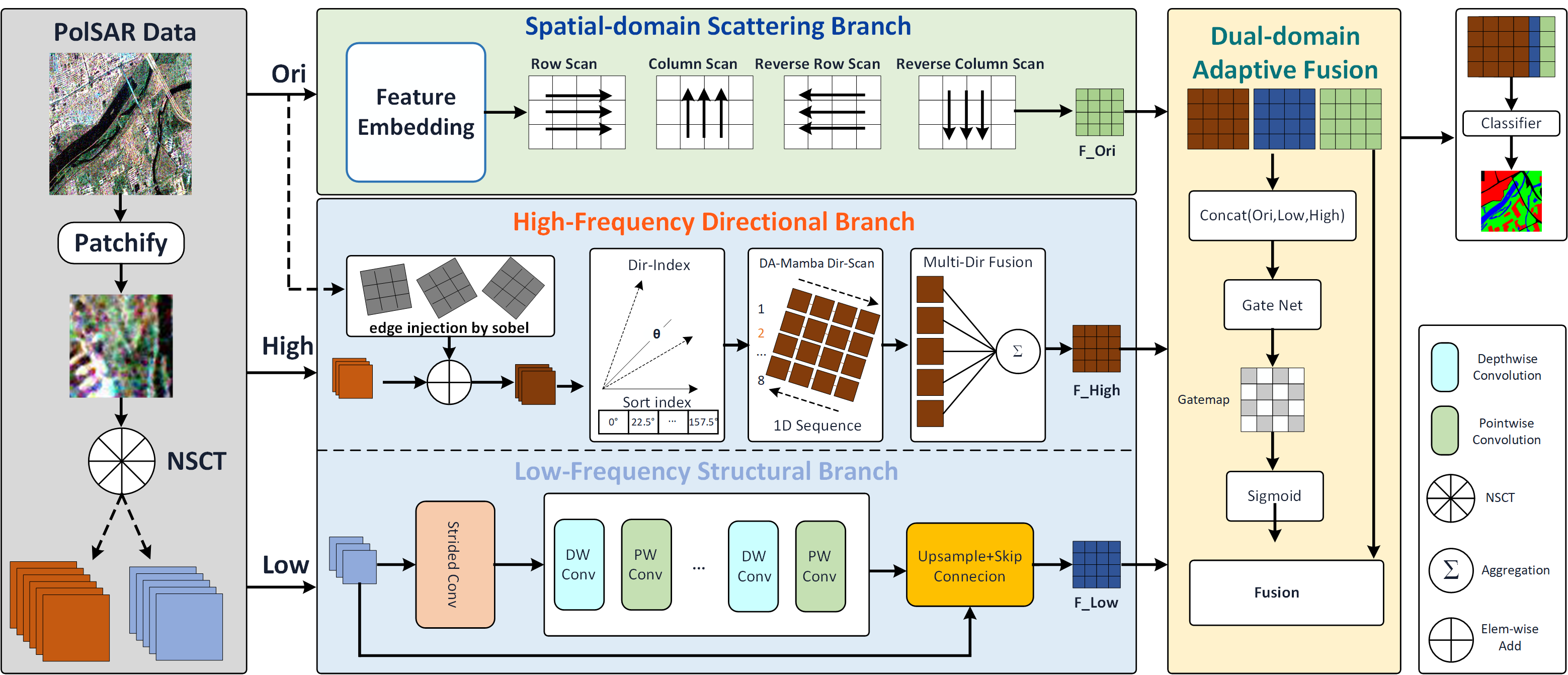} 
    \caption{Framework of the proposed method.} 
    \label{fig:fig1} 
\end{figure*}
\subsection{Network Overview}
PolSAR image classification suffers from bottlenecks in joint modeling of polarimetric scattering semantics, global spatial structures, and fine-grained directional textures. To address this issue, this work proposes DA-Mamba-DDCL, a novel direction-adaptive Mamba-based spatial-frequency dual-domain collaborative learning framework (Fig.~\ref {fig:fig1}). Unlike existing single-branch spatial learning methods, our framework builds a dual-domain modeling paradigm to exploit complementary spatial and frequency-domain cues: discriminative polarimetric scattering features, low-frequency global structural components, and high-frequency directional texture features. Moreover, a customized Direction-Adaptive Mamba (DA-Mamba) module is designed to capture diverse directional structural characteristics of PolSAR data.

In our pipeline, the Non-Subsampled Contourlet Transform (NSCT) first decomposes input PolSAR images into one low-frequency component and multi-directional high-frequency subbands. For high-frequency subbands, the DA-Mamba module models multi-orientation structural features to boost edge detection and directional texture representation. The low-frequency component is processed to extract stable global layouts, improving cross-region semantic consistency and mitigating local speckle noise. Beyond frequency-domain feature modeling, raw spatial polarimetric features are preserved to retain inherent scattering mechanisms and capture long-range spatial dependencies. Finally, a cross-branch adaptive fusion module integrates complementary dual-domain features, and the fused discriminative representations are fed into the classifier to yield final PolSAR classification results.

\subsection{DA-Mamba for Frequency Domain}


To explicitly separate structural information and muti-directional detail information in PolSAR images, we introduce the non-subsampled contourlet transform (NSCT) to decompose the original polarimetric data into low-frequency and high-frequency components. The NSCT owns shift-invariant properties and maintains full spatial resolution for all decomposed subbands.
Given the original polarimetric input $ X_{ori}\in\mathbb{R}^{C_{ori}\times H\times W}$,
the NSCT decomposition can be formulated as
\[
\mathcal{T}_{NSCT}(X_{ori})=\left\{X_{lf}, X_{hf}^{(1)}, X_{hf}^{(2)}, \ldots, X_{hf}^{(K)}\right\}
\]
where $\mathcal{T}_{NSCT}(\cdot)$ denotes the NSCT decomposition operation, $X_{lf}$ represents the low-frequency component, and $X_{hf}^{(k)}$ denotes the high-frequency component corresponding to the $k$-th directional subband. In this work, $K=8$ directional high-frequency subbands are used.
For convenience, the directional high-frequency components are stacked as
\[
X_{hf}=
\left[
X_{hf}^{(1)}, X_{hf}^{(2)}, \ldots, X_{hf}^{(K)}
\right]
\in \mathbb{R}^{K\times C_{hf}\times H\times W}.
\]
where $C_{hf}$, $H$ and $W$ denote the channel number, height and width of the feature map, respectively.

\textbf{1) Direction-adaptive Mamba for HF subbands}

\textit{(1) Edge-guided feature Enhancement}

The high-frequency(HL) components can capture fine-grained textures and boundary information of ground objects. To enhance the edge details, we propose an edge-guided HL enhancement module for preserving detailed frontier information. Given the original polarimetric image$X_{ori} \in \mathbb{R}^{C_{ori} \times H \times W}$, we extract directional edge cues through rotated Sobel filters, which can enhance directional structures and fine-grained details.
The enhancement process can be formulated as
\[
X_{hf}^{\prime} = X_{hf} + \alpha \cdot \tanh\big(\mathcal{S}(X_{ori})\big),
\]
where $\mathcal{S}(\cdot)$ denotes the edge enhancement operation implemented by rotated Sobel filters. The rotation angles are aligned with the directional decomposition angles of NSCT. $\alpha$ is a learnable parameter that controls the contribution of the original polarimetric information to the high-frequency features, and $\tanh(\cdot)$ is adopted to constrain the value range of the enhancement term.

\textit{(2) Direction-adaptive Mamba}

\begin{figure}
    \centering
    \includegraphics[width=0.45\textwidth]{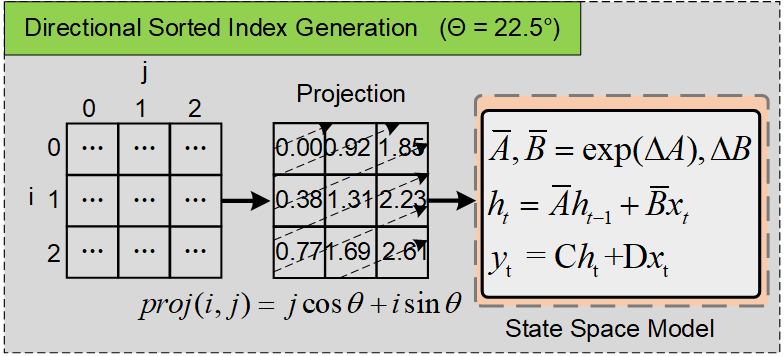}
    \caption{Illustration of the proposed direction-aware serialization in Direction-adaptive Mamba.}
    \label{fig:detail}
\end{figure}
Conventional fixed-scanning 2D Mamba neglects inter-subband directional discrepancies, limiting full utilization of NSCT’s multi-directional decomposition merits. To this end, we propose a \emph {Direction-adaptive Mamba} module to implement direction-aligned sequential modeling for oriented subbands, enhancing directional texture and fine structural feature learning. It comprises two core steps: direction-aware serialization and direction-consistent state-space modeling. The former reorganizes high-frequency subbands into orientation-matched sequences, which compute the scanning sequences along the given direction. The latter models long-range dependencies on these sequences via a shared state-space model.

\emph{Direction-aware serialization.}
As illustrated in Fig.~\ref{fig:detail}, each directional subband is first projected onto its corresponding orientation axis and sorted according to the projection values to generate a direction-aware serialization order. Let $X_{hf}^{(k)} \in \mathbb{R}^{d \times H \times W}$ represent the enhanced high-frequency feature of the $k$-th directional subband, where
$K=8$ is the total number of orientations and $d$ is the feature dimension. To align serialization with each subband’s principal direction $\theta_k$, we define a directional ordering score for spatial coordinate $(i,j)$:
\[
\zeta_k(i,j)=i\cos\theta_k+j\sin\theta_k .
\]
Sorting pixels by ascending $\zeta_k(i,j)$,yields the permutation index $\pi_k$ for direction $k$. We flatten and reorder the feature $X_{hf}^{(k)}$ into an orientation-aware sequence:
\[
S_k=\Pi_k\!\left(X_{hf}^{(k)}\right)\in\mathbb{R}^{L\times d},
\]
with $\Pi_k(\cdot)$ standing for the permutation-based serialization operation guided by $\pi_k$. This arrangement strengthens directional spatial adjacency within $S_k$, enabling the state-space model to transfer context more efficiently along the main structural orientation.

\emph{Direction-consistent state-space modeling.}
The directional sequence $S_k$ is input into a shared state-space model for long-range dependency modeling. Aligned with each subband’s principal orientation, the model propagates context along inherent structural directions with cross-branch parameter sharing. The formal formulation of this sequence modeling process is:
\[
Y_k=\mathcal{M}(S_k),
\]
where $\mathcal{M}(\cdot)$ denotes the shared Mamba state-space model.

Finally, the output sequence is restored to the original 2D spatial layout using the inverse permutation:
\[
F_{hf}^{(k)}=\Pi_k^{-1}(Y_k)\in\mathbb{R}^{d\times H\times W},
\]
where $\Pi_k^{-1}(\cdot)$ denotes the inverse operation corresponding to $\Pi_k(\cdot)$. Through the above workflow, our Direction-adaptive Mamba aligns sequence modeling with the inherent orientation of NSCT high-frequency subbands, boosting the representation of directional textures, anisotropic structures and long-range contextual dependencies.

\textit{(3) Multi-directional subbands fusion}

The Spatial Selective Fusion module(SSF) aims to boost the discriminability of fused high-frequency features via directional feature weighting, to better characterize directional discrepancies among diverse land covers. To extract the adaptive fused features, the SSF module can be expressed as
\[
F_{hf}^{fused} = \sum_{k=1}^{K} \mathcal{C}(F_{hf}^{(k)}) \odot F_{hf}^{(k)},
\]
where $\mathcal{C}(\cdot)$ denotes the convolution operation for generating directional weight maps, and $\mathrm{Softmax}(\cdot)$performs weight normalization. $\odot$ represents element-wise multiplication.

\textbf{2) Global feature learning in LF subband}

The NSCT-derived low-frequency component $X_{lf} \in \mathbb{R}^{C \times H \times W}$ retains coarse global scene structures, providing complementary semantic cues for PolSAR classification. 

For LF subband, a lightweight convolutional module is adopted to extract structural features with low computational overhead. As shown in Fig.~2, the input feature is first downsampled via strided convolution. It is then processed by stacked lightweight refinement blocks consisting of depthwise convolution, pointwise convolution, batch normalization and GELU activation. Finally, the refined feature is upsampled and fused with skip-connected input features to yield the ultimate low-frequency representation, noted by

\[
F_{low} =
\phi \left(
\operatorname{Up}
\left(
\mathcal{D}
(
\varphi_s(X_{low})
)
\right)
+
\psi(X_{low})
\right),
\]

where $\varphi_s(\cdot)$ denotes the strided projection, $\mathcal{D}(\cdot)$ represents the lightweight refinement module, $\operatorname{Up}(\cdot)$ denotes upsampling, $\psi(\cdot)$ is the skip projection, and $\phi(\cdot)$ denotes the output transformation.

\subsection{Scattering-based 2D Mamba for Spatial Domain}

Raw polarimetric data preserves full scattering characteristics and inter-channel correlations, delivering complementary spatial information to frequency-domain features for PolSAR scene understanding. Accordingly, we deploy a 2D Mamba module in the spatial domain to capture global scattering contextual patterns.

Given the original polarization input $X_{ori}\in\mathbb{R}^{C\times H\times W}$,  a shallow projection layer generates the initial feature:
\[
X_0=\mathcal{P}(X_{ori}),
\]
where $\mathcal{P}(\cdot)$  denotes feature embedding projection.

Following the 2D selective scanning paradigm,$ X_0$ is serialized along four complementary directions: forward/backward row and column axes. Each sequence is fed into a Mamba block for efficient long-range dependency modeling. The outputs are remapped to spatial space and aggregated to yield the final spatial feature:

\[
F_{ori}
=
X_0+
\frac{1}{4}
\sum_{k=1}^{4}
S_k^{-1}
\big(
\mathrm{Mamba}(S_k(X_0))
\big),
\]

where $S_k(\cdot)$ and $S_k^{-1}(\cdot)$ represent scanning and inverse scanning operations, respectively.

By combining four-directional selective scanning with state-space modeling, the original polarization branch effectively captures global contextual information while preserving intrinsic scattering characteristics, providing a strong semantic representation for subsequent feature fusion.

\subsection{Dual-domain Feature Fusion and Classification}


To fully capture the synergistic relationships among the three feature types, we first concatenate them along the channel dimension to obtain the joint feature:
\[
F_{joint} = [F_{ori}, F_{low}, F_{high}],
\]
where \(F_{ori}\), \(F_{low}\), and \(F_{high}\) represent the original polarization, low-frequency, and high-frequency feature streams, respectively.

We then adopt a gating mechanism to produce spatial weight maps for the three feature types. After normalization, the fusion weights of the three features are calculated as:
\[
W_{fusion} = \mathrm{Softmax}\left(\mathcal{G}(F_{joint})\right),
\]
where \(\mathcal{G}(\cdot)\) is the gating function used to generate the spatial weight maps, and \(\mathrm{Softmax}(\cdot)\) ensures that the weights are normalized across the three feature streams.

Thus, the weighted fusion of the three types of features can be expressed as:
\[
F_{fused} = \sum_{k=1}^{3} W_k \odot F_k,
\]
where \(W_k\) represents the fusion weight of the \(k\)-th feature stream, \(F_k\) denotes the corresponding feature stream, and \(\odot\) denotes element-wise multiplication, which is broadcasted across all channels.

This design enables the network to adaptively adjust the contribution of the three information streams for different spatial regions, thus avoiding the information imbalance caused by fixed fusion strategies. The fused feature map is then passed into the classification head for the final decision.




We adopt supervised cross-entropy loss for end-to-end training of the dual-domain collaborative network. Targeting patch-level multi-class classification, this loss aligns model predictions with ground-truth labels. The loss function is formulated as:

\[
\mathcal{L} = -\frac{1}{N} \sum_{i=1}^{N} \sum_{c=1}^{C} y_{i,c} \log(\hat{y}_{i,c}),
\]
where \(N\) is the number of samples, \(C\) is the number of classes, \(y_{i,c}\) is the ground truth label for the \(i\)-th sample and \(c\)-th class, and \(\hat{y}_{i,c}\) is the predicted probability for the \(i\)-th sample and \(c\)-th class.
\section{Experiments}

\subsection{Experimental Setup}

We evaluate the proposed method on four widely used PolSAR benchmark datasets, namely Xi'an, San Francisco, Flevoland, and Oberpfaffenhofen. Detailed descriptions of the datasets, preprocessing procedure, and visualization examples are provided in the supplementary material.

The proposed model is implemented in PyTorch and trained using the AdamW optimizer with an initial learning rate of $1\times10^{-4}$ and a weight decay of $1\times10^{-3}$. The batch size is set to 128, the patch size is 64, and all models are trained for 300 epochs. Following common PolSAR classification protocols, 5\% of labeled samples are randomly selected for training and 1\% for validation. During evaluation, Overall Accuracy (OA), Average Accuracy (AA), and the Kappa coefficient are adopted as quantitative metrics.

We compare our method with six representative state-of-the-art methods, including DFGCN~\cite{RN46}, HybridCVNet~\cite{RN38}, CV-MsAtViT~\cite{RN48}, SpectralNet~\cite{RN47}, S$^{2}$Mamba~\cite{RN19}, and NGDiffSM~\cite{RN49}. Detailed descriptions of the comparison methods and implementation settings are provided in the supplementary material.

\subsection{Results}

\begin{figure}[t]
    \centering
    \includegraphics[width=\columnwidth]{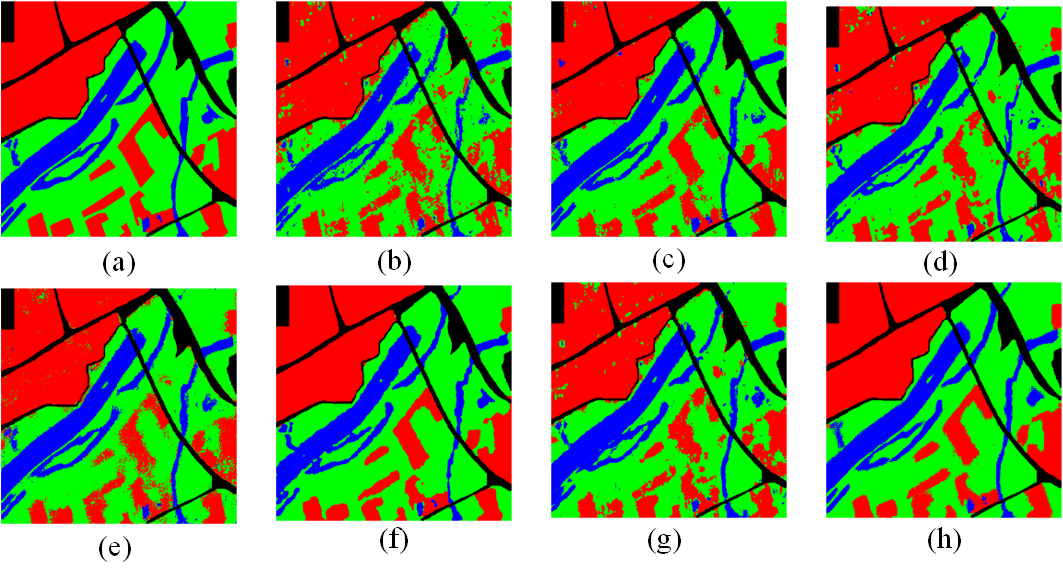}
    \caption{Qualitative comparison on the Xi'an dataset. (a) Ground Truth; (b) DFGCN; (c) HybridCVNet; (d) CVMsAtViT; (e) SpectralNet; (f) NGDiffSM; (g) S2Mamba; (h) Ours.}
    \label{fig:512_fig}
\end{figure}

\textbf{Xi'an Dataset:}
The quantitative and qualitative results on the Xi'an dataset are shown in Table~\ref{tab:512_table} and Fig.~\ref{fig:512_fig}, respectively. The proposed method achieves the best overall performance with an OA of 97.17\%, an AA of 96.64\%, and a Kappa coefficient of 95.32\%, outperforming all competing approaches. Compared with the strongest diffusion-based baseline NGDiffSM, our method further improves OA by 0.18\%. In addition, the proposed framework achieves superior performance on the water category while maintaining competitive accuracies for grass and building. As illustrated in Fig.~\ref{fig:512_fig}, our method produces cleaner classification maps with fewer isolated noisy pixels and more accurate object boundaries, demonstrating the effectiveness of contourlet-guided directional feature modeling.

\begin{figure}[t]
    \centering
    \includegraphics[width=\columnwidth]{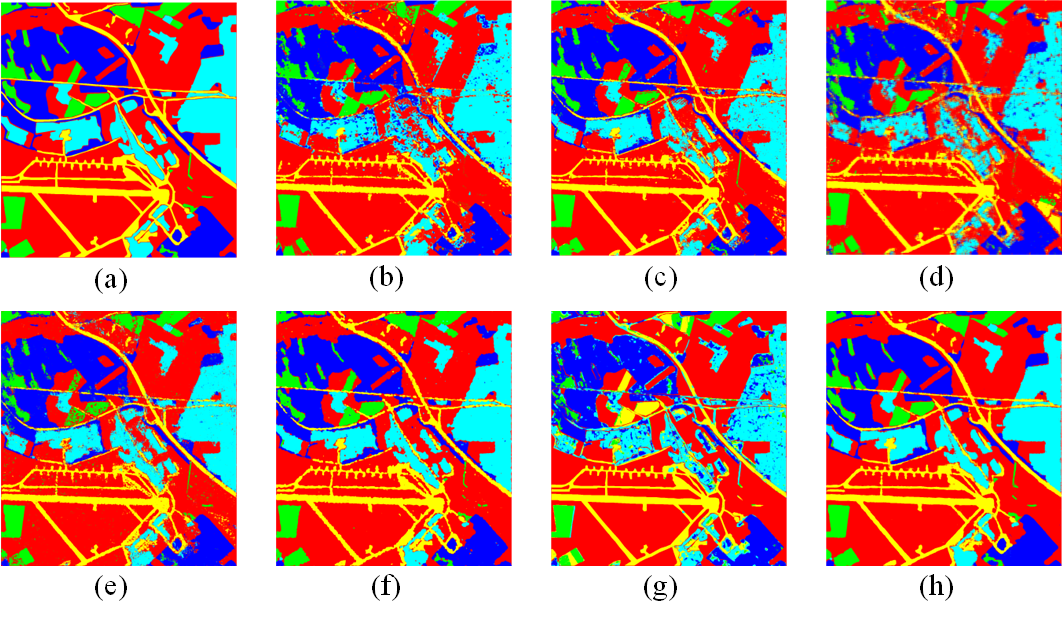}
    \caption{Qualitative comparison on the Oberpfaffenhofen dataset. (a) Ground Truth; (b) DFGCN; (c) HybridCVNet; (d) CVMsAtViT; (e) SpectralNet; (f) NGDiffSM; (g) S2Mamba; (h) Ours.}
    \label{fig:1312_fig}
\end{figure}

\textbf{Oberpfaffenhofen Dataset:}
The quantitative and qualitative comparisons on the Oberpfaffenhofen dataset are presented in Table~\ref{tab:ob_table} and Fig.~\ref{fig:1312_fig}. The proposed method obtains the highest OA, AA, and Kappa values of 95.52\%, 95.31\%, and 93.49\%, respectively, exceeding the strongest baseline NGDiffSM by 0.38\% in OA. Notably, substantial improvements are achieved for farmland and road, which exhibit complex scattering characteristics and fragmented spatial distributions. The qualitative comparison further confirms that the proposed method better preserves homogeneous regions and object boundaries while suppressing misclassified pixels. These results demonstrate the advantage of the proposed spatial-frequency collaborative framework in modeling complex PolSAR scenes.

Detailed quantitative analyses and qualitative comparisons on the San Francisco and Flevoland datasets are provided in the supplementary material.

\begin{table}[t]
\centering
\footnotesize
\caption{Classification accuracy of different methods on the Xi'an dataset (\%).}
\label{tab:512_table}

\resizebox{\columnwidth}{!}{
\begin{tabular}{l|ccccccc}
\hline
Class & DFGCN & HybridCVNet & CVMsAtViT & SpectralNet & NGDiffSM & S2Mamba & Ours \\
\hline
Water    & 84.64 & 92.01 & 93.74 & 92.82 & 92.02 & 88.89 & \textbf{94.48} \\
Grass    & 91.79 & 93.99 & 96.90 & 93.05 & \textbf{97.43} & 91.50 & 97.27 \\
Building & 87.21 & 96.72 & 90.60 & 92.96 & \textbf{98.47} & 92.71 & 98.16 \\
\hline
OA        & 89.10 & 93.50 & 94.20 & 92.98 & 96.99 & 91.53 & \textbf{97.17} \\
AA        & 87.88 & 91.70 & 93.74 & 92.94 & 95.98 & 91.03 & \textbf{96.64} \\
Kappa     & 81.84 & 89.24 & 90.38 & 88.42 & 95.02 & 86.01 & \textbf{95.32} \\
\hline
\end{tabular}
}
\end{table}

\begin{table}[t]
\centering
\caption{Classification accuracy (\%) on the Oberpfaffenhofen dataset.}
\label{tab:ob_table}
\scriptsize
\setlength{\tabcolsep}{3pt}
\renewcommand{\arraystretch}{1.05}
\resizebox{\columnwidth}{!}{
\begin{tabular}{lccccccc}
\hline
Class & DFGCN & HybridCVNet & CVMsAtViT & SpectralNet & NGDiffSM & S2Mamba & Ours \\
\hline
Bare ground & 90.51 & 94.08 & 88.21 & 93.33 & 95.20 & 91.30 & \textbf{95.71} \\
Forest      & 85.08 & 91.22 & 85.27 & 92.78 & 96.66 & 93.25 & \textbf{96.87} \\
Building    & 87.30 & 89.00 & 81.62 & 93.74 & \textbf{96.83} & 77.64 & 95.72 \\
Farmland    & 77.57 & 90.15 & 74.41 & 80.82 & 94.34 & 40.39 & \textbf{96.28} \\
Road        & 66.10 & 77.69 & 68.44 & 82.44 & 90.74 & 8.06 & \textbf{91.97} \\
\hline
OA     & 85.46 & 90.65 & 83.50 & 91.25 & 95.14 & 74.91 & \textbf{95.52} \\
AA     & 81.31 & 88.43 & 79.59 & 88.62 & 94.75 & 60.13 & \textbf{95.31} \\
Kappa  & 78.66 & 86.24 & 75.83 & 87.24 & 92.94 & 61.09 & \textbf{93.49} \\
\hline
\end{tabular}}
\end{table}

\subsection{Ablation Analysis}

\subsubsection{Ablation Study}

\begin{table*}[ht]
	\centering
	\scriptsize
	\setlength{\tabcolsep}{3.2pt}
	\caption{Ablation study of key modules on different data sets (\%).}
	\label{tab:ablation_table}
	\begin{tabular}{ccccc|ccc|ccc|ccc|ccc}
		\hline
		\multicolumn{5}{c|}{Module Configuration}
		& \multicolumn{3}{c|}{Xi'an}
		& \multicolumn{3}{c|}{San Francisco}
		& \multicolumn{3}{c|}{Oberpfaffenhofen}
		& \multicolumn{3}{c}{Flevoland} \\
		\cline{1-17}
		Ori. & Low. & High. & Ori-Guid. & Dir-Mamba
		& OA & AA & Kappa
		& OA & AA & Kappa
		& OA & AA & Kappa
		& OA & AA & Kappa \\
		\hline
		$\checkmark$ &  &  &  & 
		& 93.58 & 90.49 & 89.26
		& 99.59 & 97.67 & 99.36
		& 94.59 & 93.43 & 92.12
		& 99.94 & 99.94 & 99.92 \\

		$\checkmark$ &  & $\checkmark$ & $\checkmark$ & $\checkmark$
		& 96.39 & 95.99 & 94.05
		& 99.66 & 98.22 & 99.47
		& \textbf{95.83} & \textbf{95.82} & \textbf{93.96}
		& 99.94 & 99.95 & 99.92 \\

		$\checkmark$ & $\checkmark$ &  &  &
		& 95.45 & 94.42 & 92.47
		& \textbf{99.70} & 98.65 & \textbf{99.53}
		& 95.29 & 94.76 & 93.14
		& 99.96 & \textbf{99.97} & 99.95 \\

		$\checkmark$ & $\checkmark$ & $\checkmark$ &  & $\checkmark$
		& 96.15 & 95.86 & 93.65
		& 99.66 & 98.65 & 99.46
		& 95.62 & 95.51 & 93.64
		& 99.92 & 99.93 & 99.89 \\

		$\checkmark$ & $\checkmark$ & $\checkmark$ & $\checkmark$ &
		& 95.73 & 94.76 & 92.93
		& 99.68 & \textbf{98.79} & 99.50
		& 95.44 & 95.30 & 93.38
		& 99.93 & 99.94 & 99.90 \\

		$\checkmark$ & $\checkmark$ & $\checkmark$ & $\checkmark$ & $\checkmark$
		& \textbf{97.17} & \textbf{96.64} & \textbf{95.32}
		& 99.55 & 98.31 & 99.29
		& 95.52 & 95.31 & 93.49
		& \textbf{99.97} & \textbf{99.97} & \textbf{99.95} \\
		\hline
	\end{tabular}
\end{table*}

As shown in Table~\ref{tab:ablation_table}, using only the original polarization branch leads to limited performance, while introducing either the high-frequency or low-frequency branch improves the results in most cases. This indicates that directional texture details and low-frequency structural information provide complementary cues to the original PolSAR features.

The variant without Ori-Guided Enhancement shows performance degradation on most datasets, demonstrating that the original polarization features are beneficial for refining high-frequency representations and suppressing ambiguous directional responses. Moreover, replacing the proposed Directional Mamba with conventional 2D Mamba in the high-frequency branch leads to lower performance on Xi'an and Flevoland, suggesting that direction-aware scanning is more suitable for modeling NSCT high-frequency subbands.

Overall, the complete model achieves the best results on Xi'an and Flevoland and remains competitive on San Francisco and Oberpfaffenhofen, verifying the effectiveness of the proposed multi-branch modeling, original-feature-guided enhancement, and direction-aware high-frequency representation.

\begin{table}[htbp]
\centering
\caption{Effect of the number of decomposition directions on classification performance (\%).}
\label{tab:direction_ablation}
\begin{tabular}{lcccc}
\hline
Dataset & Directions & OA & AA & Kappa \\
\hline
Xi'an & 4  & 95.26 & 94.35 & 92.16 \\
Xi'an & 8  & \textbf{97.17} & \textbf{96.64} & \textbf{95.32} \\
Xi'an & 16 & 95.48 & 94.08 & 92.52 \\
\hline
San Francisco & 4  & \textbf{99.60} & 98.06 & \textbf{99.37} \\
San Francisco & 8  & 99.55 & 98.31 & 99.29 \\
San Francisco & 16 & 99.37 & \textbf{98.48} & 99.02 \\
\hline
Flevoland & 4  & 99.95 & 99.96 & 99.93 \\
Flevoland & 8  & \textbf{99.97} & \textbf{99.97} & \textbf{99.95} \\
Flevoland & 16 & 99.91 & 99.92 & 99.88 \\
\hline
\end{tabular}
\end{table}
\subsubsection{Effect of the number of decomposition directions}
To analyze how decomposition direction count impacts performance, we test our method with 4, 8, and 16 directional subbands. Table \ref{tab:direction_ablation} shows the
classification performance is sensitive to the directional settings. On the Xi’an dataset, OA rises from 95.26\% (4 directions) to 97.17\% (8 directions) before dropping to 95.48\% (16 directions). This implies too few directions fail to capture intricate scattering textures, while excessive directions yield redundant high-frequency signals and hinder feature fusion. The Flevoland dataset exhibits the same pattern: 8 directions yield the optimal 99.97\% OA, versus 99.95\% (4) and 99.91\% (16). The San Francisco dataset is an exception, where 4 directions deliver marginally higher OA (99.60\%) than 8 (99.55\%), with negligible performance disparity. Balancing classification accuracy and feature robustness, we fix the decomposition direction count to 8 for our model.

\subsubsection{Effect of the Mamba State Dimension}

\begin{figure}[t]
    \centering
    \includegraphics[width=0.9\columnwidth]{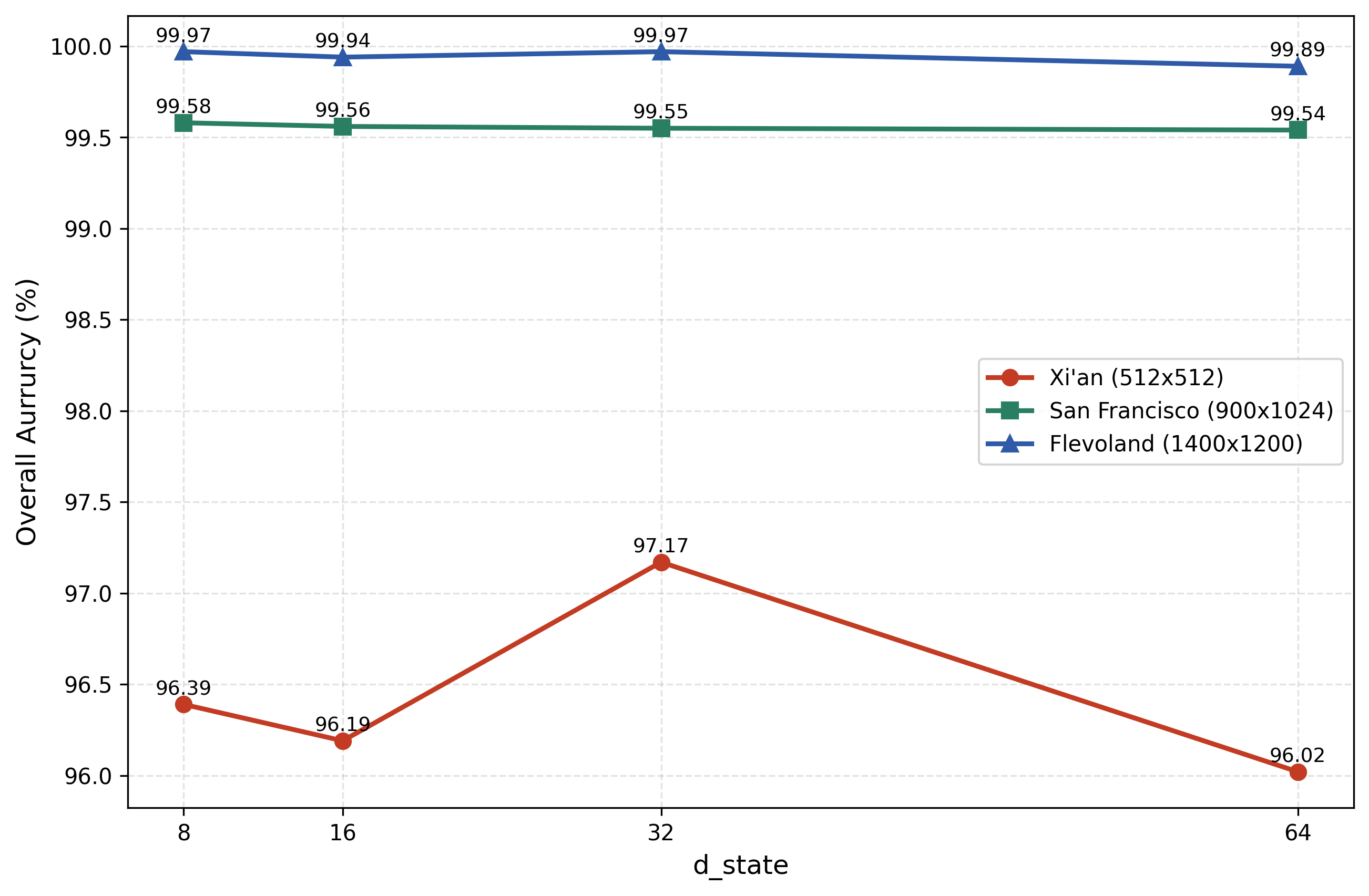}
    \caption{Effect of the Mamba state dimension on classification performance.}
    \label{fig:d_state_ablation}
\end{figure}

We evaluate the influence of the Mamba state dimension $d_{state}$ using four candidate values: 8, 16, 32, and 64. As shown in Fig.~\ref{fig:d_state_ablation}(b), a moderate state dimension generally leads to more stable performance. On the Xi'an dataset, the OA improves from 96.39\% to 97.17\% as $d_{state}$ increases from 8 to 32, but drops to 96.02\% with a state dimension of 64. This indicates that an appropriate state dimension helps improve long-range dependency modeling, whereas an excessively large state space may introduce redundant parameters and reduce generalization. Since the performance differences on San Francisco and Flevoland are relatively small, a state dimension of 32 is selected as the default setting by considering both accuracy and complexity.


\begin{figure}[t]
    \centering
    \includegraphics[width=0.9\columnwidth]{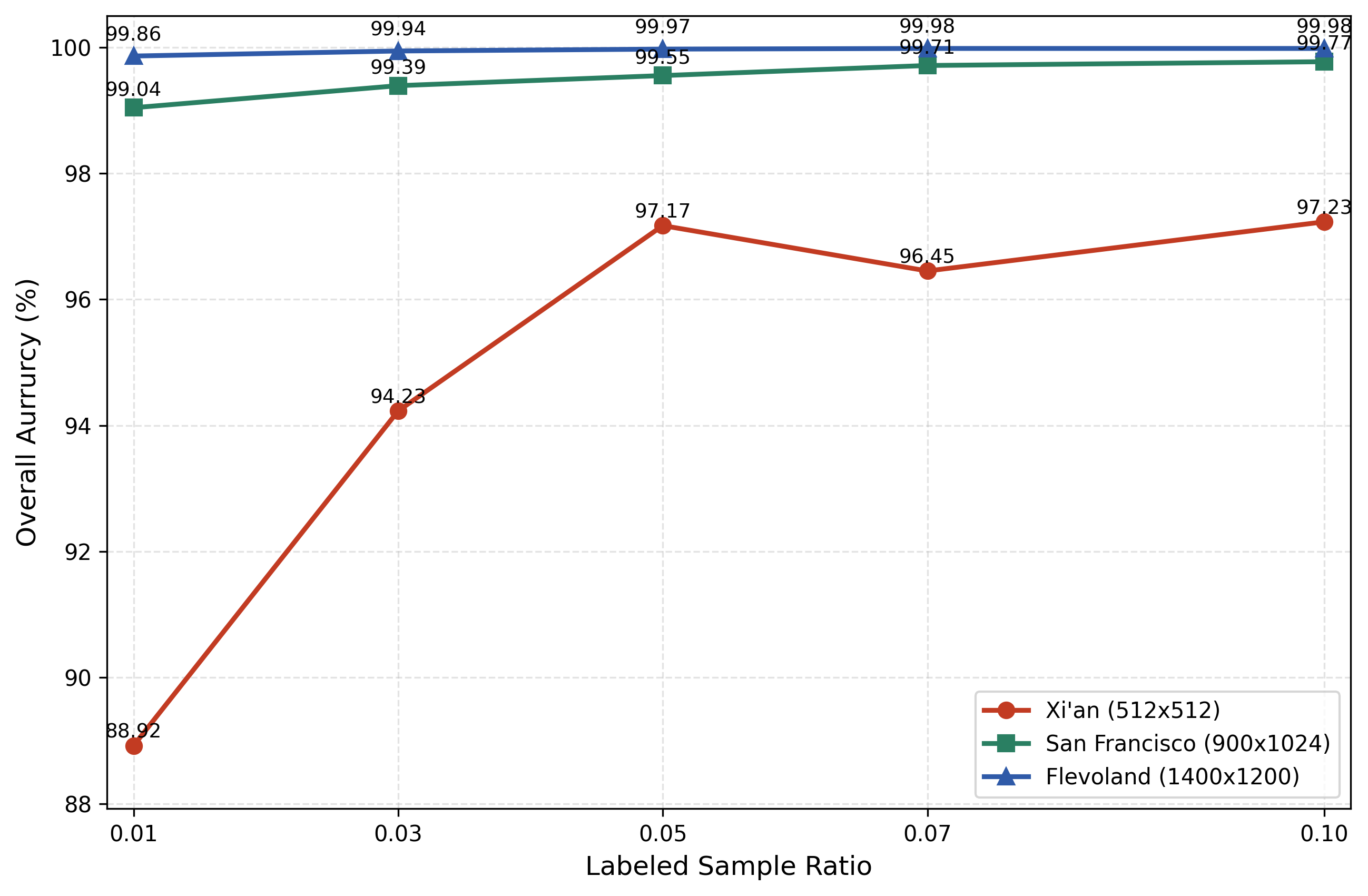}
    \caption{Effect of the training sample ratio on classification performance.}
    \label{fig:ratio_ablation}
\end{figure}

\subsubsection{Effect of Training Ratio}
We ablate training label ratios (1\%, 3\%, 5\%, 7\%, 10\%) to study how limited labeled data affects our model. Figure~\ref{fig:ratio_ablation} shows classification accuracy grows with more training labels. On the Xi’an dataset, OA jumps sharply from 88.92\% (1\%) to 97.17\% (5\%), proving extra annotations strengthen discriminative power over complex terrain. Raising the ratio further to 10\% only marginally lifts OA to 97.23\%, yielding negligible gains. San Francisco and Flevoland follow identical saturation trends, with performance plateauing under abundant labeled samples. We adopt a 5\% training split for all primary experiments: this setup enables fair comparisons against prior arts and strikes a favorable trade-off between prediction accuracy and labeling overhead.

\section{CONCLUSION}
This work proposes a dual-domain collaborative DA-Mamba-DDCL network for PolSAR land cover classification. Tailored for NSCT multi-directional decomposition, the proposed direction-adaptive Mamba module strengthens anisotropic texture learning for high-frequency subbands, and the gated dual-domain fusion adaptively aggregates raw polarimetric spatial features and frequency-domain structural cues. Experimental results verify that our approach outperforms state-of-the-art baselines in classification accuracy, speckle noise robustness and computational efficiency. However, the model suffers from limited generalization under few-sample scenarios. Future work will focus on developing semi-supervised paradigms to promote its practical generalization capability.

\bibliography{aaai2027}
\end{document}